\documentclass[prb,aps,twocolumn,showpacs]{revtex4}

\usepackage{epsfig}

\begin{document}

\title{Magnetic field dependence of the low-temperature specific heat of the electron-doped superconductor Pr$_{1.85}$Ce$_{0.15}$CuO$_{4}$}

\author{W. Yu}
  \email{weiqiang@umd.edu}
\author{B. Liang}
\author{R. L. Greene}
\affiliation{Center for Superconductivity Research, Department of Physics, University of Maryland, College Park, MD 20742}

\date{\today}

\begin{abstract} 

We remeasured the magnetic field dependence of the low-temperature specific heat of the electron-doped superconductor Pr$_{1.85}$Ce$_{0.15}$CuO$_{4 }$ 
($T_C=22\pm 2K$) under a different measurement procedure. Under field-cooling, the electronic specific heat follows $C_{el}(H, T)=\gamma (H) T$ from 
$4.5K$ down to $1.8K$. In the field range $H_{C1}<H<0.5 H_{C2}$, the Sommerfeld coefficient $\gamma (H)$ is well fit by a power-law $\gamma (H)\sim 
H^{1/2}$. This result suggests that the pairing symmetry is d-wave-like at all temperatures below $4.5K$. Our new measurement shows no evidence for 
the linear field dependence of $\gamma (H)$ found previously at $T=2K$.  

\end{abstract}
\maketitle

The pairing symmetry is an essential input to understand the mechanism of superconductivity and as well other novel properties of superconductors. It 
is generally accepted that the hole-doped ($p$-doped) high-$T_C$ cuprates have a d-wave symmetry, based on most experimental results and theoretical 
studies \cite{Harlingen_rmp_67_515, Tsuei_rmp_72_969}. However, the pairing symmetry of the corresponding electron-doped ($n$-doped) cuprates remains 
unclear. Phase-sensitive \cite{Tsuei_prl_85_182, Ariando_prl_94_167001}, ARPES \cite{Armitage_prl_86_1126, Matsui_prl_95_017003}, penetration depth 
\cite{ Kokales_prl_85_3696, Prozorov_prl_85_3700} and Raman measurements \cite{Blumberg_prl_88_107002} suggest a d-wave symmetry. However, point 
contact tunneling \cite{Biswas_prl_88_207004}, penetration depth \cite{Skinta_prl_88_207005, Kim_prl_91_097001} and specific heat 
\cite{Balci_prl_93_067001} revealed some s-wave features in the optimally-doped and overdoped samples, and led to the suggestion of a d-wave to s-wave 
symmetry transition upon doping or lowering the temperature below $4K$. Transport and ARPES measurements have shown that the Fermi surface of the 
n-doped cuprates evolves from electron-like to hole-like upon doping, with a two-carrier behavior around the optimal doping \cite{Jiang_prl_1994_1291, 
Armitage_prl_88_257001}. Recently, it was proposed that Fermi surface topology may influence the pairing symmetry which lead to a symmetry mixing 
and/or a symmetry transition upon doping \cite{Blumberg_prl_88_107002, Khodel_prb_69_144501}.

The magnetic field dependence of specific heat has been confirmed to be a valuable tool for detecting the bulk pairing symmetry, since the electronic 
specific heat $C_{el}$ in the vortex state ($H_{C1}\ll H \ll H_{C2})$ scales with H and $\sqrt{H}$ for s-wave and d-wave superconductors respectively  
\cite{Volovik_jetp_1993_469}. However, it has been pointed out that proper measurement conditions are essential to produce these theoretical 
predictions \cite{Sonier_CM_0505073}. In this paper, we report our new specific heat measurements on high-quality crystals of the n-doped 
superconductor Pr$_{1.85}$Ce$_{0.15}$CuO$_{4 }$ (PCCO). Under magnetic field cooling (FC), the field-dependence of the low temperature ($T\le 4.5K$) 
electronic specific heat has the power-law behavior $C_{el} \sim H^{0.5}$ down to $1.8K$. This d-wave-like feature is consistent with the recent phase 
sensitive tunneling \cite{Ariando_prl_94_167001} and ARPES \cite{Matsui_prl_95_017003} experiments. These results supplant our previous specific heat 
data measured on similar crystals \cite{Balci_prl_93_067001} and rule out the possibility of a temperature-dependent d-wave to s-wave symmetry 
transition at optimal doping.   

\begin{figure}
\includegraphics[width=8.5cm, height=8cm]{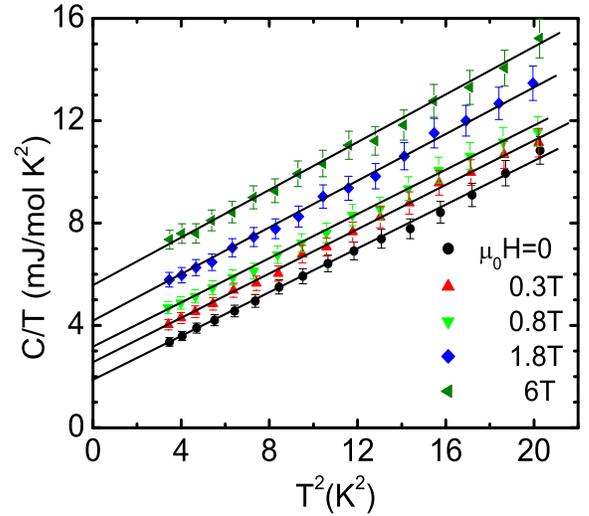}
\caption{\label{hcvst2} (color online) Total specific heat of two Pr$_{2-x}$Ce$_{x}$CuO$_{4 }$ crystals (crystal I + II) at various magnetic fields 
under field cooling conditions. The field is applied parallel to the c-axis of the crystals. Straight lines are the linear fits (see text).}
\end{figure}

The detailed sample preparation and experiment setup have been described in our previous paper \cite{Balci_prl_93_067001}. The optimally doped 
Pr$_{2-x}$Ce$_{x}$CuO$_{4 }$ ($x=0.15\pm 0.005$ determined from WDX measurements) crystals were grown by the self-flux growth technique. Here we 
selected two PCCO single crystals (crystal I and II), each with thickness of $\sim 50 \mu m$ and a mass of $\sim 2mg$. After oxygen reduction, both 
crystals show a superconducting transition at $T \sim 22K$, with a full transition width of $4K$ from SQUID magnetization measurements. The specific 
heat measurements were conducted in a Quantum Design PPMS, with a home-made specific heat puck \cite{Balci_prl_93_067001}. The magnetic field was 
applied parallel to the c-axis of the crystals. We measured separately the specific heat of crystal I, and of crystal I and II (to enhance signal to 
noise at low temperatures). As stated above, our new results were obtained under field cooling conditions. That is, for each data point with the same 
magnetic field, the field was first applied at $30K$, which is well above the superconducting transition temperature ($T_C$). The specific heat was 
then measured after cooling to a fixed temperature between $1.8K$ and $4.5K$ (see Fig.~\ref{hcvst2}). This procedure is {\it crucially  different} 
from the previous measurement \cite{Balci_prl_93_067001}, as we discuss later. Results obtained from different crystals show similar field and 
temperature dependence. Down to our lowest measured temperature ($1.8K$), no Schottky upturn was observed, which simplifies the analysis of the 
specific heat. 

In the vortex state with $ H < H_{C2}$, the low temperature specific heat of superconductors can be written as $C=C_{ph}+C_{el}$, with $C_{ph}$ and 
$C_{el} $ the phonon and the electronic contributions respectively. The phonon contribution $C_{ph}$ is represented by $\beta T^3 $ well below the 
Debye temperature. The electronic contribution  $C_{el}$ has two parts,  $C_{el} = \alpha T^2 + \gamma (H) T$, where $\alpha T^2$ is due to 
quasi-particle excitations of a d-wave superconductor ($\alpha =0$  for BCS s-wave superconductors) and $\gamma (H) T$ is due to excitations around 
the vortex cores and their extended regions \cite{Volovik_jetp_1993_469}. $\gamma (H)$ is the Sommerfeld coefficient, which varies with the magnetic 
field. In our temperature range of measurement the $\alpha T^2$ term was not detectable so the specific heat is represented by  $C=\beta T^3+\gamma 
(H) T$. The specific heat of crystal I and II from $4.5K$ to $1.8K$ is shown in Fig.~\ref{hcvst2}. For clarity, we plot $C/T$ vs $T^2$ at only a few 
magnetic fields. The straight line fits are evident for temperatures down to $1.8K$ and fields up to $6T$. We note that the slopes of these straight 
lines are about $40\%$ larger than the actual value of $\beta$ for PCCO \cite{Balci_prl_93_067001} because of a contribution from a small unknown 
amount of the addenda Wakefield thermal compound which was not subtracted from the data. We verified that the specific heat of the thermal compound 
follows $C \propto T^3$ in this temperature range. As a result, the Sommerfeld coefficient $\gamma (H)$ can be determined from the the vertical 
intercept in Fig.~\ref{hcvst2}. With increasing magnetic fields, these straight lines are shifted upwards, indicating the increase of $\gamma (H) $.

\begin{figure}
\includegraphics[width=8.5cm, height=8cm]{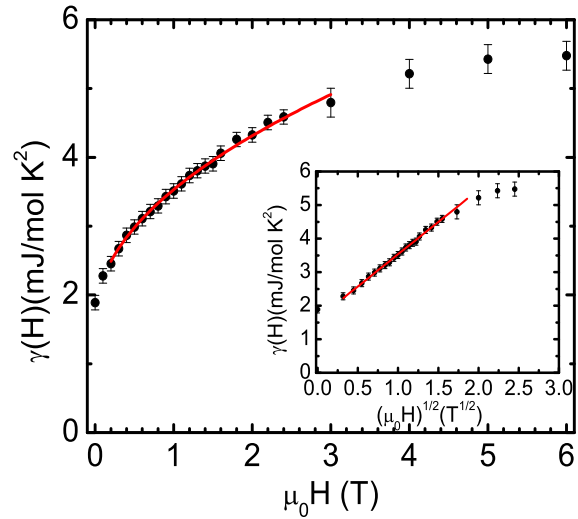}
\caption{\label{gammavsh} (color online) The field dependence of the Sommerfeld coefficient $\gamma (H)$ of Pr$_{2-x}$Ce$_{x}$CuO$_{4 }$ crystals 
(crystal I + II). The solid line is the power-law fit $\gamma (H)= \gamma ' + A H^{1/2}$ in the field range of $0.2T \le H \le 3T$. Inset: Plot of 
$\gamma (H)$ vs $(\mu _0 H)^{1/2}$; Solid line: the linear fit of $\gamma (H)= \gamma ' + A H^{1/2}$. }
\end{figure}

Fig.~\ref{gammavsh} shows the value of $\gamma (H)$ from $0T$ to $6T$. A nonzero $\gamma(H=0)$ is clearly seen, followed by a nonlinear increase of   
$\gamma (H)$ with field and a saturation behavior near $6T$.  Finite $\gamma (0)$ has been observed in many cuprate superconductors and its origin has 
been interpreted differently \cite{Hussey_ap_2002_1685, Kresin_prb_1992_6458, Phillips_prl_1990_357}. In this paper, we focus on the field-dependent 
part. Volovik pointed out that for a d-wave superconductor the quasi-particle spectrum in the extended regions of the vortex cores has a Doppler shift 
due to the supercurrent induced by the magnetic field\cite{Volovik_jetp_1993_469}. As a result, the electronic specific heat from these excitations  
follows a field dependence by $\gamma (H)= A \sqrt H $, in the range $T\ll T_C$ and $H< H_{C2}$. On the other hand, for s-wave (fully gapped) 
superconductors, the quasi-particles are thought to be confined to the vortex cores.  Then the specific heat is proportional to the magnetic field, 
$\gamma (H)= A H $, because of the linear increase of vortices with field. However, in real s-wave systems the power-law $\gamma (H)= A H ^{\alpha}$ 
with ${1/2} \le \alpha \le 1$ is frequently seen, which makes the interpretation more subtle \cite{Nakai_prb_2004_100503}. For example, for the s-wave 
superconductor Nb, a deviation from a linear field fit is clearly seen above a crossover magnetic field $H^{*}\sim 0.25H_{C2}$ 
\cite{Sonier_CM_0505073}. For $p$-doped cuprate superconductors, the $\sqrt H$ dependence has been reported for many crystals  in support of d-wave 
pairing symmetry \cite{Hussey_ap_2002_1685, Wen_CM_0508517}. However, there are also a few exceptions that need to be clarified 
\cite{Hussey_ap_2002_1685}.   

We attempt to fit our specific heat data to a d-wave power-law ($\alpha =1/2$). In the field range of $0.2T \le H \le 3T$ our data can be well fit by 
$\gamma (H)= \gamma ' + A \sqrt H $,  with $\gamma ' = 1.64 mJ/mol K^2$ and $A=1.92 \pm 0.1 mJ/mol K^2 T^{1/2}$, as shown in Fig.~\ref{gammavsh}. In 
the inset we replotted  $\gamma (H)$ against $H^{1/2}$, which shows a very good linear relation in the fitting range.  By adopting the theoretical 
form $A=\gamma _n (\frac {8a^2}{\pi H_{C2}})^{1/2}$ we obtain $\gamma _n = 4.1 mJ/mol K^2$, if we take the reasonable choices $H_{C2} \approx 6T$ and 
$a \approx 0.7$ \cite{Wang_prb_2001_094508}.

Our d-wave-like fit is supported by the following:

1) The $\sqrt H$ fit is confined to the magnetic field range of  $0.2T \le H \le 3T$. The deviation of $\gamma (H)$ from $\sqrt H$ is evident in both 
the low field ($H\le 0.1T$) and high field ($H>3T$) limit, as seen in the inset of Fig.~\ref{gammavsh}. The field range for the $\sqrt H$ behavior 
actually validates the argument that the d-wave fit only applies to the vortex state well below $H_{C2}$ \cite{Volovik_jetp_1993_469, 
Nakai_prb_2004_100503, Sonier_CM_0505073}. Thermodynamic studies show that the $H_{C1}$ is close to $0.1T$ for Pr$_{1.85}$Ce$_{0.15}$CuO$_{4 }$ 
\cite{Balci_prb_2004_140508}. The levelling off of the electronic specific heat at around $\mu _0 H=6T$ suggests that the $H_{C2}$ is about $6T$ in 
the current system \cite{Balci_prl_93_067001}. 

2) Our fitting parameter $\gamma '$ is only slightly smaller than $\gamma (0)$. It is possible that  $\gamma (0)$ is enhanced by a small trapped 
magnetic field, which is usually observed in our measurement system.

3) The high-field Sommerfeld coefficient $\gamma(H\ge H_{C2})$ is expected theoretically to be $\gamma ' + \gamma _n$ \cite{Wang_prb_2001_094508}. For 
our theoretical fit (previous paragraph), $\gamma ' + \gamma _n \sim 5.74 mJ/mol K^2 $, which is quite close to the measured value of $\gamma(H\ge 
6T)$ shown in Fig.~\ref{gammavsh}.

\begin{figure}
\includegraphics[width=8.5cm, height=8cm]{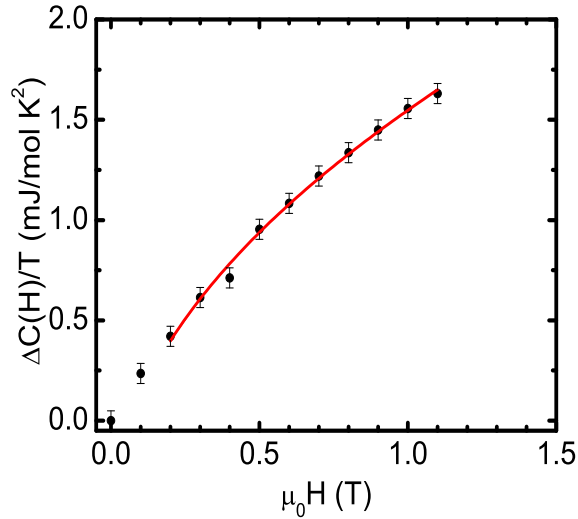}
\caption{\label{hcvsh} (color online) The field-dependence of the specific heat of Pr$_{2-x}$Ce$_{x}$CuO$_{4 }$ crystals (crystal I + II) at 
temperature $T=1.8K$. The solid line is given by $ \gamma (H)= \gamma ' + A H^{1/2}$. }
\end{figure}

In order to compare with the results in our previous paper \cite{Balci_prl_93_067001}, we also performed separate measurements and analysis at a few 
fields ($\mu _0 H \le 1.1T$) at our lowest temperature $T=1.8K $ \cite{cool_procedure}. In Fig.~\ref{hcvsh}, $\Delta C(H)/T = (C (H)- C(H=0))/T$ is 
plotted against magnetic field, by subtracting the zero-field specific heat. $\Delta C(H)/T$ is equivalent to $\Delta \gamma (H) = \gamma (H) -\gamma 
(0)$, since the phonon contribution is independent of magnetic field. The nonlinearity between $\Delta \gamma (H)$ and field is clearly seen.  A 
power-law fit $\Delta \gamma (H)= A \sqrt H $ with $A= 2.08 \pm 0.1 mJ/mol K^2 T^{1/2} $, as shown by the solid line, is reasonably good for $H\ge 
0.2T$.  
  
The power law dependence we find between the specific heat and the magnetic field at low temperatures is clearly different from our previous paper, 
where the power law exponent $\alpha$ was found to increase from $1/2$ to $1$ as the temperature decreased from $4.5K$ to $2K$ 
\cite{Balci_prl_93_067001}. The principal difference between these two works is the measurement procedure. In the previous measurement 
\cite{Balci_prl_93_067001}, the sample was cooled from above $T_{C}$ under a constant field $\sim 9T$, but then the data were taken by sweeping the 
field down at a constant temperature. In this work, we cooled the system under a fixed magnetic field from above $T_{C}$ to the measurement 
temperature. The data in Fig.~\ref{hcvst2} and Fig.~\ref{hcvsh} were obtained this way by starting from above $T_{C}$ for each field value. We found 
that the measurement procedure in our previous paper may have caused the variation of $\alpha$ for two reasons. First, the specific heat measured by 
the field sweep procedure is larger at low fields than our new data. This suggests that there is flux trapping in the sample during the field sweep 
and consequently the actual field value in the sample is larger than indicated by the PPMS. Second, below $T=4.5K$, the temperature of our heat 
capacity stage can vary during the measurement period by up to $2\%$ at $1.8K$ (decreasing to $0\%$ at $4.5K$). This could lead to an artificial 
increase of $C_{el}$ which varies with field. In our new work, the scheme of data extrapolation to $T\rightarrow 0$ (Fig.~\ref{hcvst2}) or data 
interpolation to $T\rightarrow 1.8K$ (Fig.~\ref{hcvsh}) \cite{cool_procedure}  eliminates these problems. We think that our new data measurement 
procedure is more appropriate for comparing $C_{el}(H)$ with theory. 

In summary, we remeasured the specific heat of optimal electron-doped PCCO crystals. Our new data show that the Sommerfeld coefficient, $\gamma (H)$, 
is fit nicely by $\sqrt H$ in the magnetic field range  $H_{C1}<H<0.5H_{C2}$ and the temperature range $1.8K \le  T \le 4.5K$. This result rules out  
a d-wave to s-wave symmetry transition in this temperature range for optimal-doped materials. Our result is completely consistent with the recent 
phase sensitive measurements in n-doped cuprates \cite{Ariando_prl_94_167001}. Khodel et al. proposed a possible d-wave to s- or p-wave symmetry 
transition upon doping \cite{Khodel_prb_69_144501}, due to the change of the Fermi surface topology. Further work with more over-doped crystals will 
be necessary to address this issue. 

This work is supported by the NSF under award DMR-0352735. The authors are grateful for discussions with Dr. H. Balci, S. E. Brown, Y. Dagan, and C. 
Lobb.       


\begin{thebibliography}{26}
\bibitem[{coo()}]{cool_procedure}
\bibinfo{note}{The specific data at a fixed field were collected while the
  sample temperature varied near $1.8K$. These data were then interpolated to
  achieve the data at T=$1.8K$ precisely.}
\end{thebibliography}

\end{document}